\documentclass[a4paper,twoside,12.1pt]{article}
\def\b{\bigskip }
\def\bb{\bigskip\bigskip}
\def\bbb{\bb\b}

\def\no{\noindent}
\def\r{\rightline}
\def\ce{\centerline}
\def\ve{\vfill\eject}

\def\r{\rightline}
 
 \def\g{{\got g}}
\def\L{{\cal L}}

 \def\s{\sig\mua}
\def\\mueti#1{\par\indent\llap{#1\enspace}\ignorespaces}
\def\harr#1#2{\s\muash{\\muathop{\hbox to .25 in{\rightarrowfill}}
 \li\muits^{\scriptstyle#1}_{\scriptstyle#2}}}

\def\diagra\mu#1{{\nor\muallineskip=8pt
 \nor\mualbaselineskip=0pt \\muatrix{#1}}}

\def\R{{\cal R}}

\def\today{\ifcase\month\or January\or February\or March\or April\or
May\or June\or July\or
August\or September\or October\or November\or  December\fi
\space\number\day, \number\year }

\input epsf

\r \today

\def\DD{\vec \bigtriangledown}

\def\Rr\mu{\hbox{\rm I\hskip -2pt R}}
\def\Nr\mu{\hbox{\rm I\hskip -2pt N}}
\def\Cr\mu{\hskip0.5\mu\mu \hbox{\rm l\hskip -4.5pt C\/}}
\def\w{\wedge}

\def\D{{\cal D}}

\def\p{\partial}

\def\sqr#1#2{{\vcenter{\vbox{\hrule height.#2pt
\hbox{\vrule width.#2pt height#2pt \kern#2pt
\vrule width.#2pt}
\hrule height.#2pt}}}}

 \def\square{\mathchoice\sqr34\sqr34\sqr{2.1}3\sqr{1.5}3}
\def\vac{|0\rangle}
 \def\1/2{{\scriptstyle{1\over 2}}}
 \def\a/2{{\scriptstyle{3\over 2}}}
 \def\5/2{{\scriptstyle{5\over 2}}}
 \def\7/2{{\scriptstyle{7\over 2}}}
 \def\3/4{{\scriptstyle{3\over 4}}}

\begin{document}

\def\bbb{{\cal B}}

\font\steptwo=cmb10 scaled\magstep2
\font\stepthree=cmb10 scaled\magstep4

\def\picture #1 by #2 (#3){
  \vbox to #2{
    \hrule width #1 height 0pt depth 0pt
    \vfill
    \special{picture #3} 
    }\usepackage{color}
  }

\def\scaledpicture #1 by #2 (#3 scaled #4){{
  \di\muen0=#1 \di\muen1=#2
  \divide\di\muen0 by 1000 altiply\di\muen0 by #4
  \divide\di\muen1 by 1000 altiply\di\muen1 by #4
  \picture \di\muen0 by \di\muen1 (#3 scaled #4)}
  }

\def\sqr#1#2{{\vcenter{\vbox{\hrule height.#2pt
\hbox{\vrule width.#2pt height#2pt \kern#2pt
\vrule width.#2pt}
\hrule height.#2pt}}}}

 \def\square{\mathchoice\sqr34\sqr34\sqr{2.1}3\sqr{1.5}3}
\def\vac{|0\rangle}

\def\M{{\cal M}}
\def\D{{\cal D}}
\def\L{{\cal L}}
\def\P{{\cal P}}
\def\X{{\dot{\vec X}}}
\def\N{{\cal N}}

\def\g{\Gamma}
\def\o{\omega}

\b

\ce{\bf  \huge Theory of sounds in He - II}
\bb

\ce{Christian Fronsdal } 

\b
 
\ce {\it Bhaumik Institute, Department of Physics and Astronomy}

\ce{ \it Unversity of California Los Angeles CA USA}

\b

 A dynamical model for Landau's original approach to  superfluid Helium is presented, with two velocities but only one mass density.   Second sound is an adiabatic perturbation  that involves the temperature and the  { \bf roton}, {\it aka} the  {\bf notoph.
  The action incorporates all the conservation laws, including the equation of continuity.   With only 4 canonical variables it has a  higher power of prediction than Landau's later, more complicated model, with its 8 degrees of freedom.  The roton is identified with the massless  notoph. This theory gives a very satisfactory account of second and fourth sounds.

Second  sound is  an adiabatic oscillation  of  the temperature  and both vector fields, with no net material motion.   Fourth sound  involves the roton, the temperature and  the density.

With  the experimental confirmation of gravitational waves   the relations between  Hydrodynamics and  Relativity  and particle physics have become more clear, and urgent. The appearance of the Newtonian potential in irrotational hydrodynamics comes  directly from Einstein's  equations
for the metric.     The density factor $\rho$ is essential; it is time to acknowledge the role that it plays in particle theory.

To complete the 2-vector theory we  include the massless roton mode.  Although this mode too  is  affected by the mass density, it turns out that the wave function of the unique notoph propagating mode  $\N$  satisfies the normal massless wave equation $\square \N = 0$;  the roton propagates as a free particle  in the bulk of the superfluid without meeting  resistance. In this circumstance we may have discovered the mechanism  that lies behind  the flow of He-II through very thin pores.

\bb

{email fronsdal@physics.ucla.edu} \hskip1cm webpage {fronsdal.physics.ucla.edu}

\ve

{ Table}
 
\no{ I. Introduction}  

\no{ II. The classical action principles}   

 {\it Hydrodynamics. Thermodynamics. Gauge theory. Speeds of sound.}  

\no { III. Dynamics of first and second sounds.}

{\it First sound. Second sound.
 Interpretation. }

\no{  IV. Fourth sound. }

\no { V. What comes nex?}

\bb

\ce {\bf \Large I. Introduction }

The classical action for adiabatic hydro-thermo-dynamics of  irrotational  fluid flows  allows for  the well known calculation of the speed of  sound, understood as an oscillation of the mass density and the   velocity potential at fixed, uniform  entropy (Laplace 1825) [1].   Some fluids transmit a second type of  ``sound''  that has been  interpreted as an oscillation of entropy and temperature at fixed pressure  (Tisza 1938) [2].  Experiments have confirmed that the temperature is oscillating   (Peshkov 1946) [3] and that the pressure is only weakly involved.

This paper presents an alternative interpretation of second and fourth sounds, within Landau's 2 - flow  theory [4] of phonons and rotons,  as an adiabatic oscillation of the temperature and the dynamical roton  mode, with fixed density and entropy.  The theory is an application of adiabatic thermodynamics, formulated as an action principle.    

The dynamics of the roton field ($\X$)  was identified with   the notoph (Rasetti and Regge 1972) [5], providing the  link to   Special  Relativity and Quantum Theory that is needed in any mature,  physical field theory.

 A 2-form gauge field $Y$  is related to $\X$ by $Y_{ij} = \epsilon_{ijk}X^k$.  The dynamical  roton is the massless field [7]
 
$$
\N = \rho(\DD\cdot\vec X  +  {\rm const}). \eqno(1.1)
$$

The principal new discovery  that is reported here is that {\bf second sound is an adiabatic oscillation of the temperature and $\N$}.  
\b

Section II is a brief introduction to the ideas that have led to a  dynamical formulation of Landau's theory. The speed of second sound is calculated in Section III  and fourth sound is tackled in Section IV.


\bb

\ce {\bf \Large II. The classical action principles}
\b
  
\ce {\bf   \large Hydrodynamics}

The essence of classical hydrodynamics is expressed by two equations,
the equation of continuity,
$$
\dot \rho  + \DD\cdot \rho\vec v = 0                 \eqno(2.1)
$$
and the Bernoulli equation
$$
{\p \over \p t} \vec v = -  \DD\vec v^2/2 - {1\over \rho}\DD p - \DD\varphi.     \eqno(2.2)
$$ 
Here $\rho$ is the (mass) density and $p$ is the pressure.
It applies only to irrotational  flows, when the velocity takes the form $\vec v = - \DD\Phi.$  The two equations of motion are the Euler - Lagrange equations of a classical  action principle.   The field $\varphi$ is the Newtonian potential.\footnote {This theory is what remains of a relativistic theory  when the relativistic scalar $\psi$ is expanded as  $\psi = c^2t + \Phi +  O(1/c^2)$ and, $g_{00} = c^2 + 2\phi + O(1/c^2)$ and the other components are Lorenzian.}

For some of the most elementary flows   another branch of hydrodynamics  must be invoked. In a popular,  didactic experiment a glass of water is placed on a turntable. After some time the water is seen to be turning with the glass  like a solid body, the surface rising towards the edge to form a meniscus.  In the theory that is used to explain this phenomenon  the velocity is a time derivative, $\X$, and the `Bernoulli equation' takes a different form,
 $$
 \DD \X^2/2  - \DD\varphi -  {1\over \rho}\DD p  = 0.      \eqno(2.3)
$$ 
This theory,   by itself,  is not an alternative to the irrotational theory.  It does not have an equation of continuity; instead $\X$ is subject to constraints, as expected of a vector field. The inclusion  of  the Newtonian potential in this equation is {\it ad hoc}, it  can not be justified by an application of General Relativity  and  the vector field $\X$ is   not affected by the transformations of the Galilei group. In conclusion, we need both types of vector fields to explain some of the simplest experiments.

The study of lementary applications like these are incontrovertible evidence that two kinds of flow are needed in hydrodynamis.  A satisfactory description of the waterglass - on - turntable and the whorls seen in the wake of ships  was proposed by Onsager.  (Onsager 1962) [6]. 

Both theories  can be expressed as action principles; the irrotational Lagrangian  density is
$$
\L_1[\rho,\Phi,\varphi] = \rho(\dot\Phi - \DD\Phi^2/2 -\varphi) - W_1[\rho]     \eqno(2.4)
$$
and Eq. (2.3) - without $\varphi$ - is the Euler-Lagrange equation of
$$
\L_2[\rho,\vec X] =   \rho(\DD \X^2/2 ) - W_2[\rho].\eqno(2.5)
$$
The density factor is traditional in $\L_1$,  less so in $\L_2$; its appearance in both  is crucial.  \footnote { That compressibilty of air is what makes flight possible was 
understood by Leonardo da Vinci in the 15th century.}

Current hydrodynamics  results from adding (2.4) and (2.5),
$$
\L_{Hydro}[\rho.\Phi, \vec X] = \L_1[\rho,\Phi] + \L_2[\rho, \vec X] +  {\kappa\rho \over 2} d\psi dY .\eqno(2.6)
$$
The last term will be explained below.

 The idea of two independent vector fields  was already introduced by Landau  [4] in his theory of  superfluid Helium,  his phonon and roton velocities fields  are $ - \DD\Phi$ and $\X$.  

The classical theory of ordinary sound is derived  from 
  the Lagrangian (2.6),  
  $$
\L[\rho. \Phi, \vec X]  = \rho(\dot\Phi - K - \varphi) - W[\rho],)\eqno(2.7)
$$  
with the kinetic potential 
$$
K =  (\DD\phi)^2 /2 - (\DD \X)^2/2 - \kappa \X\cdot \DD\Phi. \eqno(2.8)
$$

 This Lagrangian is invariant under the transformations of the Galilei group.  (The  field $\vec X$ is inert, up to a change of gauge.)  The flow $\rho (\kappa \X - \DD\Phi)$ is identified  by the property of being conserved, as expressed by equation of continuity, derived from the Lagrangian by variation of $\Phi$.  
\b

As we shall show, this is a suitable action for Landau's phonons  and rotons  and a wide range of other applications of adiabatic hydro-thermo-dynamics. In the
literature inspired by Landau's work on superfluids one finds that the applications make little use of roton  dynamics; instead the field $\X$ is more or less fixed.   
 It is, therefore,  not surprising  to find that the theory, in its original, non - relativistic context,  is characterized by strong constraints, as has been revealed by completion of the theory (Rasetti and Regge 1972). This is what brings the number of independent variables of  hydrodynamics down to just 4.

The completed roton theory is a relativistic  gauge theory. Like electrodynamics, it was completed with its development as a quantized gauge theory  (Ogievetskij and Polubarinov 1963) [7] and Green, Schwartz and Witten (1987) [9].  Both relativity and quantum theory are needed for the formulation 
 of unitarity. We  return to this topic below. 
 
 Both $\L_1$ and $\L_2$ are non-relativistic limits of relativistic  field theories; the former is a limit of
 $$
{1\over 2}\rho(g^{\mu\nu}\psi_{,\mu}\psi_{,\nu} - c^2)  - W[\rho].\eqno(2.9).
$$ 
The non-relativistic limit includes the Newtonian potential defined by 
$$
g_{00} = c^2t + 2\varphi + O(1/c^2),~~~  \psi = c^2+ \Phi + O(1/c^2)\eqno(2.10)
$$
This is the origin of the Newtonian potential in Eq.(2.4) [9]. 
Its appearence in (2.5) cannot be justified by General Relativity.  
\ve

\ce{\bf  \large  Thermodynamics}

The use of a variational principle for  (adiabatic)  thermodynamics is not often seen in the literature.  There follows a resum\'e that shows that  the basic equations are  the Euler-Lagrange equations of a simple action. This reformulation of adiabatic thermodynamics  contains nothing  that is unfamiliar.  What it does is to set the limits of the applications;  \footnote { It is evidently  incompatible with an oscillating entropy. } it fixes the Hamiltonian, the kinetic potential and the angular momentum and it puts us in a better position to confront new applications, such as gravitational waves [11] and the speed of second sound.

The equations that define adiabatic thermodynamics of  a uniform system at rest are
$$
{\p F(T,V) \over \p V} + P =  0,~~~{\p F(T,\rho) \over \p T} + S = 0, \eqno(2.11)   
$$
where $V$ is the volume, $F$ is the Helmholtz free energy and  $P$ is the 
pressure. We prefer to formulate the theory in terms of densities, 
$$
s = \rho S,~~~ f(T,\rho) = \rho F(T,V).
$$
 Following Callen  we set the local version of Eq.s (2.11)   
$$
\rho{\p f\over \p \rho} - f = p, ~~~ 
{\p f\over \p T} + s = 0.                        \eqno
$$

 Consider the action
 $$
 A_1[ \Phi, \rho,T, S,P]  = \int dt\bigg ( \int_\Sigma \L_1 -  \int_{\p \Sigma } P\bigg).                         \eqno(2.12)
 $$
 Here $P$ is the 3-form of pressure on the multifaceted boundary. 
 
 The Lagrangian density is                                      
 $$
\L_1 =  \rho(\dot \Phi - \DD \Phi^2/2 -\varphi)- f(T,\rho)-s T.          \eqno(2.13)          
 $$
  Assume  that the 
specific entropy density  $S$ is fixed, constant and uniform.  Vary this Lagrangian with respect to local variations of $\rho$ and $ T$,  with $S$,  $P$ and - temporarily -  the boundary $\p \Sigma$, fixed, then the Euler - Lagrange equations are as follows.

  Variation of  $A_1$ with respect to $\Phi$ gives the equation of continuity, with $ \vec v =  -\DD\Phi$;
 $$
\dot\rho + \DD\cdot (\rho\vec v) = 0, \eqno(2.14)
$$

\no Variation with respect to  $T$ gives the adiabatic relation:
 $$
 {\p \over \p T} f + s = 0;            \eqno(2.15)
 $$
 it can be used to eliminate the temperature.  
 \b

 \no{\bf Theorem.}  When $ s = \rho S$, $S$ fixed, constant and uniform,
 then
 $$
\DD {\p \over  \p \rho}(f + sT) = {1\over\rho}\DD p.\eqno(2.16)
$$     
 Local variation  of $\L_1$, Eq.(2.13), by $\rho$, followed by the elimination of $T$,  leads to  the Bernoulli equation in the original form, Eq. (2.2).
 
$$
 \DD{\dot \Phi }- \DD(\DD \Phi)^2/2 - \DD\varphi - {1\over \rho}\DD p = 0.      
 $$
     There is  a proof  in Fronsdal (2020) [10].
 \b

 Finally, variation of  the boundary gives 
 $$
\L_1 |_{\p \Sigma}   =  P. \eqno(2.17)
$$
 On-shell,  \underbar{on the boundary},
 $$
 p  =    \rho {\p f  \over \p\rho} - f = \L_1  -  \rho {\p \L_1  \over \p\rho} 
 =   P. \eqno(2.18)
 $$
 The first equality agrees with the first of Eq.s (2.11), but since it  is 
 taken to hold in a wider context  it may be regarded as a definition; the second is  a consequence of the fact that  $ - f$ is the only term in the Lagrangian density that is not linear in $\rho$. The last equality confirms the identification of $p$ as the pressure, an extrapolation of $P$ from the boundary to the 
interior. 
 
  We shall replace $\L_1$ by $\L_1 + \L_2$, as in Eq. (2.6).
\bb

\ce{\bf Gauge theory}

The gauge theory behind the roton field $\X$ was discovered by Rasetti and Regge [5]. It is the theory of a massless 2-form, components  $(Y_{\mu\nu})$. The  free action density is  $\rho dY^2$.  For  hydrodynamics  the complete action density is 
$$
\L_2[\rho, Y] = \sqrt{-g}
{c^2\over 12}\rho \,dY^2 + {\kappa\over 2}\rho\, dY d\psi;\eqno(2.19)
$$
The 2-form $Y_{\mu\nu}$ is related to $\vec X$ and $\psi$ is related to $\Phi$ by (2.10). 
The non relativistic Lagrangian in Eq. (2.12) is derived from  $\L_1$ in Eq. (2.9) and   $ \L_2$ is a limit of $\L[\rho,Y]$ in (2.19).

1. The field  $\psi$ is a scalar field with a vacuum expectation value, $\psi = \Phi + c^2 t$. The field $\Phi$ transforms, together with the velocity $ - \DD\Phi$,  under the Galilei group, in the usual way, making $\L_1$ invariant under this group.  

2. The components of the 2-form are
$$
Y_{ij} = \epsilon_{ijk}X^k,~~~Y_{0i} =: \eta_i.\eqno(2.20)
$$
The vector field $\vec \eta$ is a gauge field, variation of the 
action with respect to $\vec \eta$ gives  the constraint - the gauge condition -
$$
\DD \w \vec m = 0,~~~ \vec m :=  \rho(\X + \kappa
\DD\Phi),\eqno(2.21)
$$
with the general solution
$$
\vec m = -\DD \tau.
$$
 A special choice for the  gauge parameter  $\tau$   is required for a massless mode to be recognized. This mode is
$$
\N := \rho(\DD\cdot \vec X + \kappa).\eqno(2.22)
$$
It is the only   propagating field of this gauge theory. The free field equation is (Ref.s  [11]).
$$
\square \,\N = 0.    \eqno(2.23)
$$
{\bf Remark.} Non-relativistic electrodynamics, a limit   when the speed of light - $c$ -  tends to infinity, makes sense only in the absence of the magnetic field. And non-relativistic hydrodynamics is the regime where   $\N = 0$, for this field, like $\vec B$, enters the Lagrangian and the equations of motion multiplied by $c^2$.  Consequently, any study of a  steady configuration  will be one in which $\N$ is negligible. 
\bb

\ce {\bf \Large III. Dynamics of first and second sound}

\ce{\bf First sound}

 The classical theory of (first) sound propagation rests on the Eulerian theory,
with the Lagrangian density  $\L_1$ and the two equations of motion (2.1) and (2.2).  
The speed of propagation  is expressed in terms of the adiabatic derivative of the pressure 
$$
C_1 = \sqrt{d p(\rho,S)\over d \rho}.                            \eqno(3.1)
$$

This equation is used in the quoted sources (Arp {\it et al} [12], Brooks and Donnelly [13], Maynard  [14] )  to determine the equation of state. We shall obtain  similar formulas for   second and fourth sounds.

First sound is an oscillation of the density and the velocity potential,
$\rho$ and $ \Phi $,    and $S$,  fixed.  The speed of first (ordinary) sound is usually calculated for a  plane wave, a  first order perturbation of a static configuration with uniform density.  The two Euler - Lagrange equations are
$$
\dot \rho - \DD\cdot \rho \DD\Phi = 0,~~~~
\dot \Phi - {\p (f + s T)\over\p \rho}\bigg|_T = 0.\eqno(3.2)
$$ 
Eliminating $T$, or using Eq. (2.16) and differentiations  leads, in first order perturbation theory,  to
$$
\ddot \rho = \rho \DD\cdot\dot{\vec v}(\rho, S),~~~~
 ~~~~\DD\cdot\dot{\vec v} =  {1\over\rho}{\p p(\rho, S)\over \p \rho}\bigg|_S \Delta\rho  \eqno(3.3)
$$ 
and to (3.1). Note: in this case $\vec v = -\DD\Phi$, since $\X = 0$.
 
\b

\ce{\bf \large Second sound }

Since the notoph is a massless particle, we   add a Stefan-Boltzmann  term to  the  enternal energy density: 
$$
u(T,\rho)  \to  \tilde u (T,\rho, \N) = u + {\alpha\over  k} T^k\ N. ~~~\N :=  \rho(\DD\cdot\X + \kappa). \eqno(3.4)
$$

 Let $(T_0,\rho_0, \vec X_0, \N_0)$ be a stationary  solution of the Euler - Lagrange equations for the Lagrangian
$$
\L = \rho(\dot\Phi - K)-f-sT- {\alpha\over k} \N T^k
$$
$$
K = - \beta \X^2/2 + \vec v^2/2,~~~~\beta = 1 +\kappa^2; \eqno(3.5)
$$
this is an alternative expression for Eq.(2.6)  and $\rho\vec v = \rho(\kappa\X - \DD\Phi $ is the conserved current.

Second sound is a first order,  adiabatic  deformation
$$
(T_0, \X_0)\to 
(T_0 + dT, \vec X_0 + d\vec X),
$$
with $ \N_0 = 0,    \vec v_0 = 0$ and $dp = 0, d\rho = 0, d\vec v = 0$.


\b

To the experimenter  second sound is excited by a forced oscillation of the temperature at the boundary and  that is not registered by a pressure sensitive  microphone, hence $dp \approx 0$.  

We must review the equations that govern these oscillations. 
\b

\hrule\b

1.  The relevant part of the internal energy density is 
$$
-f -sT -  {\alpha\over k} \N T^k + \rho\beta\X^2/2
$$
In adiabatic thermodynamics, for any fixed value of S, the theory is an isolated  Lagrangian action principle and $ u $  is the Hamiltonian density. For any adiabatic variation    the integrated internal energy  is at a minimum.

  Variation  with respect to $T$ and $\vec X$,  $\rho$ fixed and uniform, gives the two Euler - Lagrange equations. From variation of $T$: 
$$
\int dT{\p( \tilde f\ + sT)\over  dT}\bigg|_{S,\rho,\N}  
 = \int  dT\bigg(\rho{\p (\tilde  F + ST)\over \p T}\bigg|_{S,\rho}  
 + {\alpha\over k} T^k {d\N\over dT} \bigg)
    \eqno(3.6)
    $$
    In the first order of the perturbation this quantity is zero,
    $$ 
d(\rho C_V)   -\alpha T^{k -1}d\N = 0.
    \eqno(3.7)
$$
Explanation: The  internal energy density is $\tilde u$ and this has  values that are measured and that give the recorded values of $C_V$; but
in the cited papers $\N$ does not represent  another variable; it is just   a function of $T$, so  their $\rho C_V$  includes a term $- (\alpha/k)T^k(d\N/dT)$.   
 
Derivation with respect to the time gives
$$
\rho {\p C_V \over \p T} \dot T - \alpha T^{k-1} d\dot \N = 0.
$$
This is valid to first order in perturbation theory if $\N_0 = 0$, as is natural under the circumstances.   Under the same conditions,
$$
\rho{\p C_V\over \p T}\ddot T - \alpha T^{k-1}
 \ddot \N = 0.  \eqno(3.8)
$$

 \hrule\b

  2.   From variation of  $\vec X$,
 
$$
\rho \beta d\X\cdot \X - {\alpha\over k} T^k d\N 
= 0,  
$$
or
$$
d\vec X \cdot (\beta \,d\ddot{\vec {X}}  -  \alpha  T^{k-1} \DD T) = 0.    \eqno(3.9)
$$

 From (3.7) and the divergence of  (3.8) follows that  
$$
\rho{\p C_V\over \p T}\bigg|_p  \ddot T - \alpha  T^{k-1}d\ddot \N = 0,~~~~ 
 \beta d\ddot \N  -  \alpha\rho   T^{k-1}\Delta T = 0  \eqno(3.9) 
$$
 and the speed  $C_2$ is  
$$
 C_2  = {\alpha\over \sqrt \beta}   T^{k-1}  {\bigg ({\p C_V \over \p T}\bigg|_p}\bigg)^{-1/2}.\eqno(3.10)
$$

The values  of $\p C_V/\p T|_p$  will be taken from experimental data, for $0< p < 25 MPa$.
\bb

\ce{\bf\large  Direct comparison with experimental values}

 Arp {\it et al} [12], Brooks and Donnelly [13] and Maynard [14] have collected results from many experiments. Their results for $C_V$ are plotted  in Fig.1, along with our very simple interpolation.\footnote {Our interpolation formula was needed in the lowest interval of temperature only,   there was no need for the elaborate interpolation use
d by Arp.} The logarithmic singularity was placed  on the $\lambda$ line. 

Our interpolation for $C_V$ is, for $p = 0$, $0.4 < T < 2.4$
$$
C_V =  
-\ln((2.18 - T)^{5/2} + 10^{-30}) + 3.35 - 4.5 T + 1.6 T^2, ~~~.\eqno(3.11)
$$
Units are joules/gram.

\epsfxsize.5\vsize
\centerline{\epsfbox{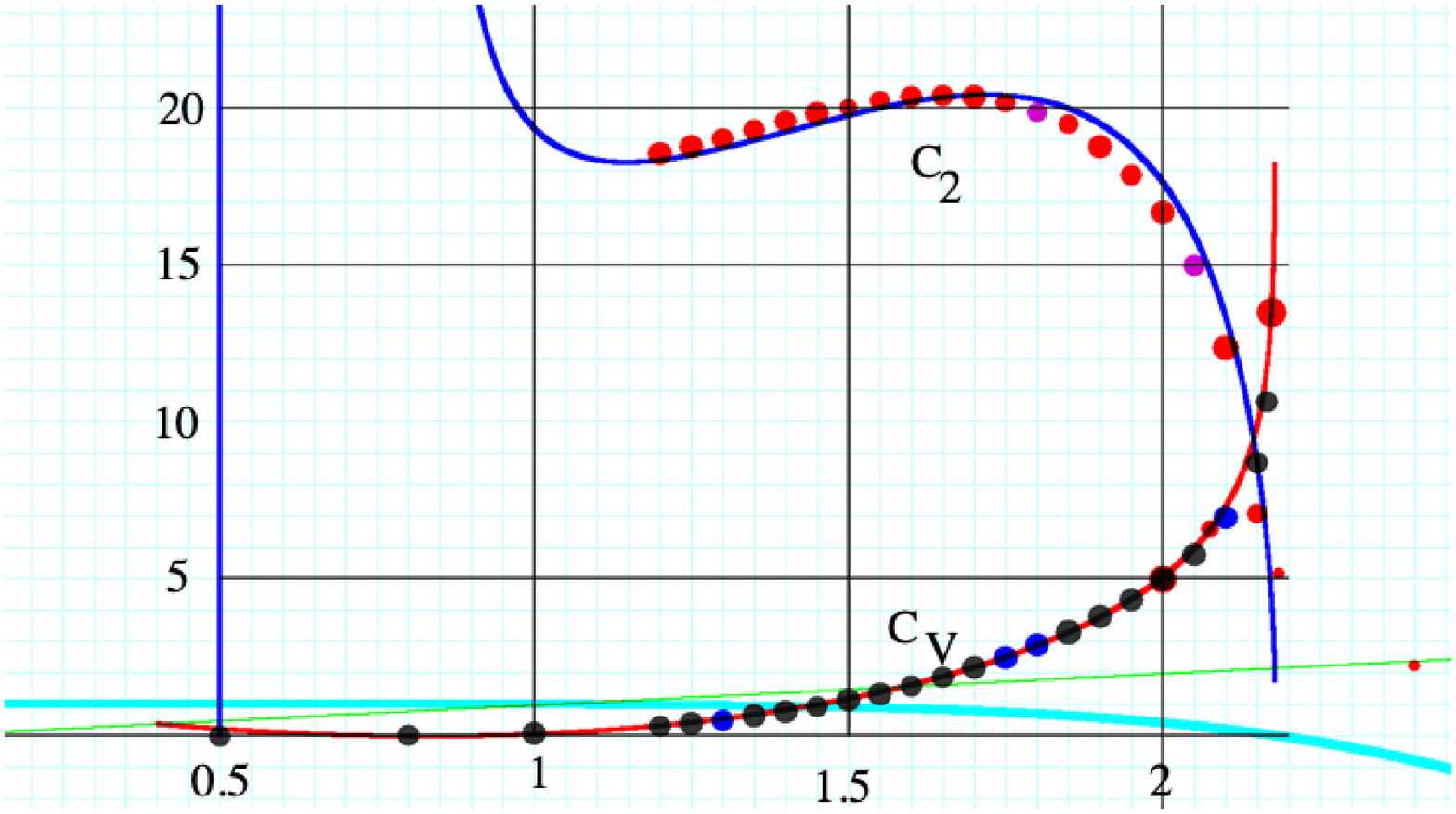}}
 \parindent=1pc

{\small Fig.1. The lower part shows values of $C_V$ (in joules) determined  by measurements. The solid curve is our simple interpolation of the data. 
This interpolation was used to calculate the speed $C_2$ of second sound, 
using Eq. (3.13), showed for $k= 3~ (units~ $m/s$)$; it is our prediction for the speed of second sound in He-II. }
\b

The curves $C_V(T)$ and $C_2(T)$ are shown in Fig. 1.  The lowest value of $C_V$ on the interpolation curve is -.033 at $T$ = .8282. The
  calculation stops at $T = .7476$, near the point where the experimenters loose their signal (Williams and Rosenbaum 1979)  [15] and peaks at $T$= 2.18046.   Only the overall factor, 17.5 in Eq. (3.16)   could be  adjusted for a best fit.
  
   Similar fits were obtained for $p  = 2MPa$  after a small adjustment of the parameters:
   $$
C_V = - \ln((2.165 - T)^{2.5} + 10^{-30})  + 4.0  - 5.85 x + 2.1 x^2.\eqno(3.12)
$$
The minimum of the interpolation curve is .00389 at $T$ = .700.  The curve begins  at $T$ = .8287,  and peaks at $T = 2.17875$,
at the $\lambda$ line. 

Calculations have verified similar agreement for pressure 5, 10, 15, 20 and 25 MPa.

\b

{
\b
 
Given the experimental data for the values of $C_V$,   the theory predicts the speed of second  sound to be given up to a multiplicative constant by 
 Eq. (3.16). These formulas, with $k = 3$,  give very good fits from the $\lambda $ line down to $T = .8$, where $C_V$ has a local minimum and the signal is lost. Fitting the overall constant factor to the experiment we find that, for $p = 0$ and for $p=2$,  \footnote  {  In the quoted reviews velocities are given in m/sec, energy densities in joules. } 
  the final result is
 $$
 C_2 =  {\alpha\over \sqrt \beta}~T^2\bigg({\p C_V \over\p T}\bigg)^{-1/2},~~~.8 < T < T_\lambda. \eqno(3.13)
 $$
 with ${\alpha/\sqrt \beta} = 17.5 $ in the units m/s and joules. 
 Here $ C_V$ was taken from the tables in terms of joules and the velocites were expressed in terms of  $m/s$.  

 \bb
\  

\ce{ \bf \Large Interpretation}

In the term $(\alpha/k) T^k \N $  that we have included in the internal energy, $\N$ is the notoph amplitude. The power $k$ in $T^k$ was left open to be determined by measurements. The experimental value is $k = 3$. The  factor $T^3$ is proportional to the number of quanta  predicted by Planck's theory. The new term in the internal energy density is thus identified as  the Stefan - Boltzmann term associated with the notoph.   
\bb

Why include  $\N  T^3$ instead of $aT^4$.  The notoph is a new experience and the wisest course is to accept the value provided by the experiments, which fixes the value of $K$ at 3..

We have set the parameter $\N_0$ equal to zero. This is because the experiments were made in vessels of a size such that  boundary effects, the origin of capillary effects, are expected to be weak. The effect of varying this parameter away from zero is insignificant. 
$$
$$$$
$$

\bb

\ce{\bf \large IV. Fourth sound} 

Fourth sound is observed in thin films  and in containers packed with silicon wafers. The usual interpretation is that the ``normal component" remains at rest; we shall assume that  
$$
 \vec v_0 = 0,~~~\DD\Phi_1 = 0,
$$
that $S$ is fixed, constant and uniform, while $\vec X, \rho$ and $T$ oscillate together. Of the four equations of motion these three are relevant for the determination of the speed;
\b

\hrule\b

1.  Equation of continuity:
$$
d\dot\rho + \kappa  \rho d(\DD\cdot \X) = 0.
$$
 To zero order in perturbation theory the system is taken  to be stationary , and both terms are  zero.  For simplicity we shall replace $\vec X $ by $\N$ as the independent variable. Consider a plane wave perturbation,  
 then to first order  the equation reduces to 
$$
\dot\rho_1 +\kappa \rho_1 (\DD\cdot \X_0) + \kappa \rho_0 (\DD\cdot \X_1) = 0. \eqno(4.1)
$$
It is clearly vital to know something about the zeroth approximation.  

In the bulk of the fluid the field $ \X_0 $ is stationary and $(\DD\cdot \X_0)$ is expected to vanish; in that case, in first order of perturbation,  
$$
d\rho  + \kappa d\N  =  \rho_0,~~~{\rm constant};
$$ 
this will allow us to eliminate the density from the Bernoulli equation.  

\b

\hrule\b

2.  The adiabatic condition:
$$
  {\p( \tilde f + s T) \over \p T} \bigg|_{\rho,\N,S}dT =  
 \rho{ \p( \tilde F +ST \over \p T}\bigg|_S dT 
 $$
 $$
  =  -  {\alpha\over 3} T^3  \N - {\p( \tilde f + s T) \over \p \rho} \bigg|_{\,T\N,S}d\rho 
$$
   and using Eq. (2.16):
   $$
     {\p( \tilde f + s T) \over \p \rho} \bigg|_{T,\N,S} = {1\over \rho}{\p p\over \p\rho}
 $$
 to get
$$
  {\p( \tilde f + s T) \over \p T} \bigg|_{\rho,\N,S}dT   
  =  -  {\alpha\over 3} T^3  d\N
  - {\p p\over \p \rho} d \rho,    
  $$
  or
  $$
dC_V   = ( \alpha T^2 + \kappa 
  C_1^2  )d\N - \rho_0C_1^2
$$
 The  time derivatives:
  $$
  {\p C_V\over \p T}\dot T +  \rho_0 {\p C_1^2 \over \p T}\dot T  = ( \alpha T^2 + 
  \kappa  C_1^2) \dot N 
    $$ 
 
   \b
    
    \hrule\b
   
   3. The Bernoulli equation  is,   
   $$
   d \ddot{\vec X} = \alpha  T^{k-1} \DD T,    \eqno(4.3)
$$
or
 $$
 \ddot \N  -  \alpha T^2 \Delta T = 0,
  $$

 and together they give
 $$
 {C_4}^2 = {\ddot T\over \Delta T}
 = 
 {\ddot T \over \ddot \N}
  {\ddot\N\over \Delta T}
= 
{\bigg({\alpha\over 100} T^2  +  \kappa   {C_1}^2\bigg) \over  {1\over 10}{\p C_V\over \p T}   + \rho_0 {\p\over \p T}C_1^2 }{{\alpha T^2\over 100} }.\eqno (4.4)
 $$\
The unit of velocity is here $10^4cm/sec,$ and  $C_V$ is in joules. In the numerator $\alpha = 17.5\beta$ as in the calculation of $C_2$ and $1/100$ converts $\alpha$ to the new unit of speed. The factor 1/10 in the denominator
is valid when $C_V,$ is expressed in joules, as  taken from the tables.

\epsfxsize.4\vsize
\centerline{\epsfbox{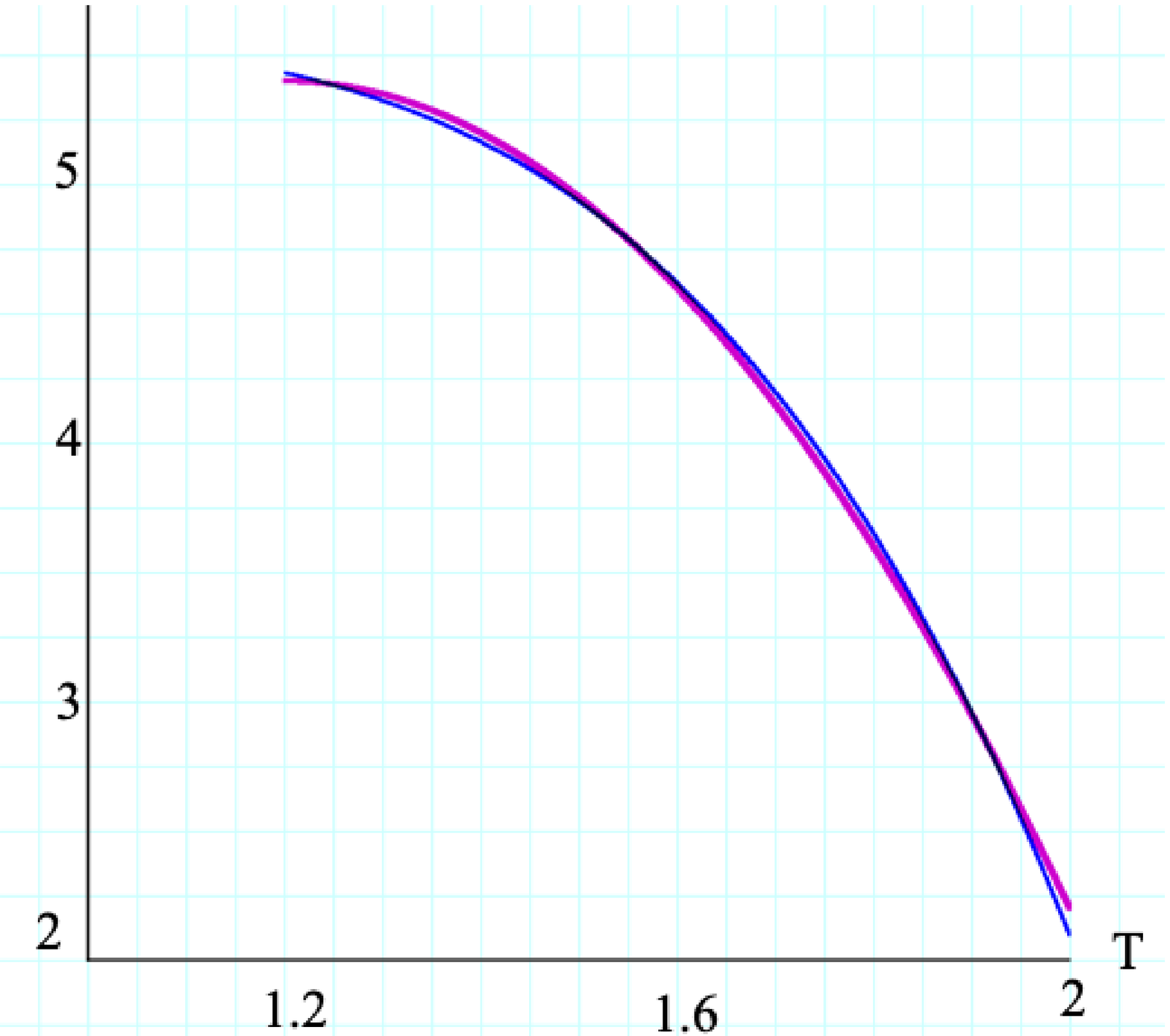}}
 \parindent=1pc
 
\small{Fig.2.  The relation (4.4).  Red line: Interpolation by the author of measurements of $C_4$. Blue line: Values of $C_4$
calculated from experimental values of $C_1, C_2$ and $C_V$ 
 reported in ref.s [13-15].}
 \epsfxsize.5
 \vsize
 \parindent=1pc
\b

Fig. 2 summarizes the result for $C_4$.   The red  line is the square of  fourth sound,  interpolated  from the experimental data [13-15].  It is  almost covered by the blue line, the value given by Eq. (4.4). The fit was made with only one free parameter, the physical parameter $\kappa$ of the fluid, for the first time determined experimentally. The vertical coordinate is the square of the speed  of fourth sound in units of $(10^4 cm/sec)^2$.
\bb


 The value of the parameter $\alpha/\sqrt \beta$ was determined in Section III. That leaves  the value of  $\kappa$ as the sole free parameter; the value  determined by  using the earlier value of $\alpha$ is
 $$
 \kappa = .556 \pm .0005.
 $$
If instead  Eq. (4.4) is used to determine both parameters  then 
$\alpha/\sqrt \beta$ is bracketed between 17.0 and 18.0.

That gives a unique theory, with no free parameters, that can be used to  predict the strength of capillary effects  and other properties of He - II. 

The analytic interpolations used for $C_1$ and $C_4$ were 
$$
{C_1}^2 = 5.63+0.05*[1.2-x]-(0.77*[x-1.2]^2)
$$
$$
{C_4}^2 = 54000-(50000*[x-1.2]^2)
$$
\ve

\ce{\Large \bf V. What comes next?}
 
This paper reports another application  of a version of Landau's
2-vector theory of superfluids. It should be pointed out that the
alternative idea of two densities has found no direct experimental support.
The number ``$\rho_s/\rho_n$'' is fixed in terms of $\rho, T$, and $p$;
it is not an independent variable. [13]

The need for  extra variables, besides a velocity potential,  a density and the temperature,  was demonstrated at the end of the 17th century
and yet the first viable suggestion in that direction was Landau's idea of two velocities, at first  in a very narrow context.

The two `versions' of hydrodynamics date from the beginning. They have been said to be equivalent, but that is evidently not the case.
The rotons are strongly associated with the socalled `Lagrangian version' of hydrodynamics. This was pointed out in an important paper by Rasetti and Regge, but  that paper had repercussions in string theory only.  [16] 

Today we see the roton-notoph identification  as the coming-of age of hydrodynamics, with applications to a  large class of fluid phenomena, including   capillary action, flight and gravitational waves. The recently confirmed unification of General Relativity  with Particle Theory has  given  new impetus to  bringing hydrodynamics into contact with both.  
We hope that the present paper will stimulate a more unified approach to these important branches of physics, by showing that hydrodynamics can be approached by methods that have been proper to Particle Physics, and profit from it. The approach assumes a precise model of fluids and  makes detailed predictions on its own, without special adaptations in each special case.  

To end where we began, superfluids still pose challenges.
The existence of spin is obvious but details need to be examined.
The spectacular ability of He-II to penetrate very fine pores is probably a manifestation of capillary phenomenon, related to the properties of the massless notoph, but this needs to be clarified. 
Finally, the growing importance of notoph = roton makes it urgent to discover  how to detect it, in the CMB and in the laboratory.
\bb

Computation codes are available on request from the author.
\bb

{\Large  \bf Acknowledgements}

I thank Gary Williams for discussions and information,
and  Joe Rudnick for conversations. I also wish to
acknowledge the crucial reference to the paper [5],
by Alexander Zheltukhin. I also thank  Chair David Saltzberg for support.

 \epsfxsize.6
 \vsize
 \parindent=1pc

 \epsfxsize.8\hsize
\vskip.5cm
\ve

  {\bf \large References}
\b

 \no [1] Laplace, P.S., {\it Trait\'e de M\'echanique},  
 Duprat, Paris (1825)
 
\no \no [2] Tisza, L., {\it Transport Phenomena in Helium II},  

Nature volume 141, page 913 (1938)

\no [3] Peshkov, V.,   ``Second Sound in Helium II.

 J. of Physics, {\bf  8}, 381-389.'' Scientific Research (1944)

\no [4] Landau, L., ``Theory of Superfluidity in 

Helium II", Phys.Rev. {\bf 60} 356-358 (1941)


\no [5] Rasetti, M. and Regge, T., ``Quantum vortices 
and diff (R3)'', in Lecture Notes in Physics, 

\no [6] Onsager, L. Private conversation, 1962

Springer-Verlag. (1962)

\no [7] Ogievetskij, V.I.  and  Polubarinov, I.V., 

``The notoph and its possible interactions'',
    
    Lecture Notes in Physics, 
    Springer-Verlag 
    173  (1962) 

\no [8] Green, M.,  Schwartz, J.  and Witten, E.,

``{Superstrings'', Princeton U. Press  (1987) 

 \no [9]    Fronsdal, C. ``Ideal stars in General Relativity'', 

Gen. Rel. Grav. {\bf 39}, 1971-2000  (2007)  

\no  [10] Callen, H. {\it Thermodynamics}, Wiley (1960)

\no [11] Fronsdal, C., ``Hydronamic sources for 

Gravitational Waves''. (2020) 

\no   [12]  Arp. V.D.,   McCarty,  R.D. and Friend,  D.F.,

 ``Thermophysical Properties of Helium-4 from 0.8 
 
 to 1500 K with Pressures to 2000 MPa'',

Physical and Chemical Properties Division

Chemical Science and Technology Laboratory,

National Institute of Standards and Technology

325 Broadway Boulder, Colorado 80303-3328 (1998)

\no [13] Brooks, J.S. and  Donnelly, R.J., 

``The calculated thermodynamic properties of 

superfluid Helium'', J.Phys.Chem.Ref.Data {\bf 6} 51 ((1977)

\no  [14] Maynard, J.,  ``Determination of the

 thermodynamics of He II from sound-velocity data. 
 
 The dynamics of Helium - II from the speeds of
 
  sound'' m  Phys. Rev.  B {\bf 14} 1976 - 3891 (1976)

 
\no [15] Williams, G. A.  and  Rosenbaum, R.,

 ``Fifth sound in superfluid  $ {~}^4 He$ below  $1 K$'', 
 
  Phys. Rev. B.  {\bf 20} 4738 - 4740   (1979)

\end{document}